%% file: main.tex
\documentclass{acm_proc_article-sp}

\usepackage{amssymb,latexsym,amsfonts,amsmath,stmaryrd}
\usepackage{graphicx}
\usepackage{alltt}
\usepackage{subfigure}
\usepackage{pritam}
\usepackage{macros}
\usepackage{algorithm}
\usepackage{algorithmic}

\usepackage{eso-pic}
\usepackage{graphicx}
\usepackage{color}

\newdef{definition}{Definition}
\newdef{lemma}{Lemma}
\newtheorem{theorem}{Theorem}
\newtheorem{proposition}{Proposition}

\newcommand{\R}{{\mathbb{R}}}

\newcommand{\N}{{\mathbb{N}}}

\newcommand{\Post}{\mathrm{Post}}

\newcommand{\U}{\mathsf{U}}
\newcommand{\W}{\mathsf{W}}
\newcommand{\w}{\mathsf{w}}
\newcommand{\x}{\mathsf{x}}

\newcommand{\z}{\mathsf{z}}

\renewcommand{\d}{\mathbf{d}}

\newcommand{\rTo}{\im}

\newcommand{\tool}{{\sc PessoaLTL}}
\begin{document}

\renewcommand{\r}{\textcolor{red}} 
\renewcommand{\b}{\textcolor{blue}} 
\def\mynote#1{{\sf $\clubsuit$ \textcolor{red}{ #1 }$\clubsuit$}}

\title{Safety-Guarantee Controller Synthesis\\for Cyber-Physical Systems}
%
%
%
%
%

\numberofauthors{3} 
%

%
%
\author{
\alignauthor Pritam Roy\\
\affaddr{UC Los Angeles}   \\
\email{pritam@ee.ucla.edu}
\alignauthor Paulo Tabuada\\
\affaddr{UC Los Angeles}   \\
\email{tabuada@ee.ucla.edu}
\alignauthor Rupak Majumdar\\
\affaddr{MPI-SWS and UCLA}   \\
\email{rupak@cs.ucla.edu}}

\date{30 July 1999}

\maketitle
\begin{abstract}
The verification and validation of cyber-physical systems is known to be a difficult problem due to the different 
modeling abstractions used for control components and for software components. A recent trend to address this 
difficulty is to reduce the need for verification by adopting correct-by-design methodologies. According to the 
correct-by-design paradigm, one seeks to automatically synthesize a controller that can be refined into code and 
that enforces temporal specifications on the cyber-physical system. In this paper we consider an instance of this 
problem where the specifications are given by a fragment of Linear Temporal Logic (LTL) and the physical environment is 
described by a smooth differential equation. 
The contribution of this paper is to show that synthesis for cyber-physical systems is viable 
by considering a fragment of LTL that is expressive enough to describe interesting properties but 
simple enough to avoid Safra's construction. 
We report on two examples 
illustrating a preliminary implementation of these techniques on the tool {\tool}.
\end{abstract}

%
%

\input{intro}
\input{background}

\input{safe-ltl}
\input{refinement}
\input{casestudy}

\section*{Acknowledgement}
We like to thank Manuel Mazo jr, for patiently answering our numerous quaries on the tool PESSOA.
\bibliographystyle{abbrv}
\bibliography{main}  
%
%

\end{document}

%% file: intro.tex
\section{Introduction}

The correct-by-design, or {\em controller synthesis}, paradigm offers a compelling alternative to current 
system design methodologies relying on extensive testing and/or verification to prove correctness. 
Intuitively, synthesis is the problem of algorithmically constructing an implementation from a given
specification of the desired functionality and performance, and a partial model of the system.
Controller synthesis has been studied in various forms in different communities, differing in the form of the model
and the specification. 
For example, in continuous control theory, the partial model is the open
loop plant
\[
\dot{x} = f(x,u),
\]
and the controller is a feedback function $u = k(x)$ such that the controlled
system $\dot{x} = f(x,k(x))$ satisfies certain stability
and performance criteria.
Similarly, in (discrete) {\em reactive synthesis}, the partial implementation is usually an
input-enabled, unconstrained automaton, the specification is given
as a temporal logic formula capturing the good behaviors of the system, 
and the controller is an automaton ensuring that its product with the partial
implementation only generates good behaviors.

Over the past decades, there has been a convergence of control-theoretic methods
with automata-theoretic ones, in order to model {\em hybrid} or {\em cyber-physical}
systems in which discrete components interact with continuous ones.
These systems are often complex yet safety-critical, and thus, the application of
program synthesis techniques ---as opposed to the current practice of design and
extensive verification and validation--- is likely to have a large impact.
However, there are some key technical challenges that have to be overcome in order
to apply synthesis to this domain.

First, we have to abstract the underlying continuous state space into discrete
parts so that reactive synthesis techniques can be applied. 
Moreover, such abstractions need to be constructed in such a way 
that a controller designed for the abstraction can be refined to a controller 
enforcing the specification on the original continuous model.

Second, the specification language must be expressive enough to capture many
properties of interest in the domain.
In the reactive synthesis world, {\em linear temporal logic} (LTL) \cite{Pnueli77}
(or equivalently, automata over infinite words \cite{VW94}) is usually considered
as a robust and expressive specification formalism.
Synthesis algorithms based on deep automata-theoretic constructions 
\cite{BL69,Rabin69,McNaughton,PR89a,PR89b,KV05} are well-known
for this formalism.
Unfortunately, these algorithms have very high theoretical and practical complexities. Theoretically, the problem is complete for 2EXPTIME.
Moreover, Safra's determinization construction \cite{Safra}, 
a key step in the algorithms, is extremely difficult to implement, and the best
implementations so far can only handle small automata.
This has limited the possibility of practical synthesis tools.

In this paper, we present \tool, an automatic synthesis tool for cyber-physical systems.
\tool\ takes as input a controlled differential equation modeling the physical components, 
a specification consisting of two parts: a safety part in safe-LTL and an easily determinizable liveness part,
and a parameter $\varepsilon$ specifying the desired \emph{precision}, and outputs, if possible,
a software controller that ensures that the model together with the controller
satisfies the specification up to precision $\varepsilon$ (in a technical sense).
The controller is refined to Simulink blocks for closed-loop simulation.

We overcome the two challenges mentioned above in the following way.
First, we use recent techniques reported in~\cite{PGT08,ZPT10,MazoDT10} 
to compute discrete abstractions of the differential equation model of the underlying continuous state space.
Second, we use a restricted subset of LTL for our specification language,
chosen to be expressive enough to naturally capture many requirements that frequently arise in cyber-physical
systems design, and yet enabling controller synthesis without Safra's construction (or the manipulation
of co-B\"uchi tree automata \cite{KV05}).

Our choice of the specification formalism is driven by our 
observation that many specifications for controller synthesis problems in embedded systems and robotics
essentially consist of 
an ``involved'' safety part (stating that the system should always remain in ``safe'' states) and 
a ``simple'' liveness
or guarantee part (stating that eventually the system should reach a special set of states). 
For example, a typical requirement in robotic applications is to reach a goal state while avoiding
obstacles. 
A typical problem in control is to force a system to move between different operating points while staying within 
a desired operational envelope. 
This occurs, e.g., when we press a button in an elevator requesting that we reach a different floor while 
maintaining the elevator velocity and acceleration within certain limits for safety as well as comfort reasons.
Accordingly, our specification language consists of two parts: a safety part in safe LTL, and a guarantee
part given as an until formula. 
We use the fact that automata for safe LTL can be determinized using the usual subset construction \cite{KV01},
letting us avoid Safra's construction in the implementation.
Moreover, we can symbolically compute maximal strategies for the safety part.
In a second step, we can compute the strategy to ensure the guarantee part while ensuring the safety specification.
Although our synthesis algorithms are based on enforcing a safety invariant on the product of the system
and the automaton constructed from the safe LTL formula, the use of safe LTL directly allows us to write
specifications more naturally than if using invariants.

We developed \tool~as an extension of {\sc Pessoa}\footnote{Available from
\texttt{http://www.cyphylab.ee.ucla.edu/pessoa}.
}
using both the abstraction algorithms as well as a solver for safety games using BDDs provided by {\sc Pessoa}.
We report preliminary results on the use of \tool.  
Drawing inspiration from robotics, we illustrate by two nontrivial 
examples how embedded control software synthesis problems can be automatically solved. 
The first example considers the motion planning problem with obstacles and requires a LTL 
formula comprising both safety as well as guarantee properties. 
In the second example we consider a more detailed model for the robot by incorporating information 
about the protocol used to mediate between the sensors and the main processor. 
Since the main processor mail fail to acquire sensor measurements, we consider the 
requirement of reducing the robot velocity, or even completely stopping 
the robot, when not enough measurements are acquired. 
While in the worst case, the complexity of the algorithm is still 2EXPTIME \cite{KV01}, in practice, the 
subset construction has not been a bottleneck.

\smallskip
\noindent\textbf{Related work}
We have already mentioned the rich history of reactive synthesis using automata-theoretic techniques.
Work on the synthesis problem for cyber-physical systems is quite recent. 
The use of finite-state abstractions of differential equations and hybrid systems 
to solve synthesis problems has been pursued by several authors~\cite{ReachQuant,KB08,Reissig09,KGFP09,Stormed,Tarraf10}. 
However, no new novel synthesis algorithms, at the automata level, are proposed in these references. 

Most tools for synthesis restrict speicifications to state invariants.
This is mostly because automata theoretic synthesis algorithms for general
LTL properties require a complex determinization step \cite{Safra} which is 
hard to implement efficiently \cite{AlthoffTW05,TasiranHB95}.

In~\cite{LTLC,WTM10} controller synthesis enforcing temporal requirements on cyber-physical systems 
is discussed.
Although different synthesis algorithms are proposed in these references, both assume 
a bounded temporal horizon for the satisfaction of the property. 
The work~\cite{LTLC} uses model checking algorithms to find the feasible set of inputs. 
These inputs are bounded, since it is based on bounded temporal horizon assumptions.
The liveness properties with bounded horizon are examples of bounded-safe properties. 
The fragment of LTL handled by \tool~includes all bounded-safe properties. Furthermore, 
\tool~also supports guarantee properties that require no restrictions on the time it takes for satisfaction.

In~\cite{Jobstm07b,Jobstm07c}, the authors have also restricted attention to specification formalisms
which have efficient game solving algorithms, and used such algorithms to synthesize hardware components.
Our focus here is embedded and robotics applications, for which our restricted specification language
is a good fit. The abstraction of differential equation models for the physical components is an added dimension
of complexity in our case. 
%

The synthesis of switching policies for cyber-physical systems is discussed 
in~\cite{JGGT10}. 
Although, the resulting switching policies enforce the desired specifications, 
the work in~\cite{JGGT10} assumes that the continuous dynamics in each mode is fixed. 
In contrast, our algorithms do not assume the a priory existence of different modes with different dynamics.

While our constructions do not introduce any new deep insight into the nature of synthesis, we believe our specification
formalism and implemented algorithms represent a practical sweet spot in controller synthesis for cyber-physical systems.

%% file: background.tex
\section{Background}

\subsection{Systems}

We consider the following notion of system that will be used to model
software components as well as the abstraction of physical 
components.

\begin{definition}
\label{Def:System}
A system 
$$S=(X,X_0,U,\rTo,Y,H)$$ 
consists of:
a set of states $X$;
a set of initial states $X_0\subseteq X$;
a set of inputs $U$;
a transition relation $\rTo\subseteq X\times U\times X$;
a set of outputs $Y$;
and
an output map $H:X\to Y$.
\end{definition}

A system is said to be {\em finite} when the set of states $X$ is finite. 
When the set of outputs $Y$ of a system 
$S$ is equipped with a metric $\d:Y\times Y\to\R_0^+$, we say that $S$ is a metric system. Metric systems will be used to 
formalize finite abstractions of differential equations in Section~\ref{SSec:AB}.

We write $x\rTo^u x'$ when $(x,u,x')\in \rTo$.
For such a transition, state $x'$ is called a \mbox{{\em $u$-successor}}, or simply 
{\em successor}, of state $x$. 
Similarly, $x$ is called a {\em $u$-predecessor}, or {\em predecessor}, of state 
$x'$.  
For technical reasons, we assume that for every $x$ and $u$, there is some $x'$ such that $x\rTo^u x'$.
We denote the set of $u$-successors of a state $x$ by $\Post_u(x)$. 
A system is said to be {\em deterministic} if $(x,u,x') \in \rTo$ and $(x,u,x'') \in \rTo$ implies $x'=x''$,
or equivalently, if $\Post_u(x)$ is a singleton for each $x\in X$ and $i\in U$.

A {\em run} of a system $S$ is an infinite sequence
\begin{equation}\label{eq-run}
x_0 \rTo^{u_0} x_1 \rTo^{u_1} \ldots
\end{equation}
where $x_0\in X_0$, and for each $i\geq 0$, we have $x_i \rTo^{u_i} x_{i+1}$.
The {\em outputs} associated with the run \eqref{eq-run} is the trace
$$H(x_0)H(x_1)\ldots \in Y^\omega.$$


Given an infinite string $\mathsf{z}\in Z^\omega$, we will use the notation 
$\mathsf{z}(i)$ to denote the $i$th element in the string $\mathsf{z}$ and 
the notation $\mathsf{z}[k]$ to denote the infinite string obtained 
from $\mathsf{z}$ by removing its first $k$ elements, i.e., $\mathsf{z}[k](i)=\mathsf{z}(i+k)$.

The notion of system in Definition~\ref{Def:System} allows for nondeterminism 
in the sense that for a given state $x\in X$ and input $u\in U$, 
there may be more than one $u$-successor of $x$. 
We assume that once the input $u$ is chosen at the state $x$, 
the exact $u$-successor of $x$ is selected from $\Post_u(x)$ by the environment. 
We regard this nondeterminism as the adversarial influence of the environment, and consider
a two-person game between the controller (player~0) and the non-determinism (player~1).
%

\subsection{Controllers}

A {\em strategy} for the controller (player 0) in a system $S=(X,X_0,U,\rTo,Y,H)$ 
is a mapping $\pi_0 : (X \times U)^* \times X \mapsto U$ that associates with every 
non-empty finite sequence of states and inputs ending in $X$, representing the 
past history of the game, an action. 
A {\em strategy} for player 1 is a mapping $\pi_1 : (X \times U)^* \times X \times U \mapsto X$ 
that associates with every non-empty finite sequence of states and inputs ending in $x \in X$ 
and after action $u \in U$ has been taken, representing the past history of the game, 
a successor state $x' \in Post_u(x)$. 
A controller strategy $\pi_0$ is \emph{memoryless} 
if the strategy depends on the current state only 
i.e., $\forall x\in X,\ \forall \mathsf{z}, \mathsf{w} \in (X \times U)^*,\ \pi_0(\mathsf{z}\cdot x) = \pi_0(\mathsf{w}\cdot x)$.

An initial state $x_0 \in X_0$, strategy $\pi_0$ for player $0$, and $\pi_1$ for player $1$ uniquely determine a run: 
\begin{equation}
\label{Outcome}
Outcome(x_0,\pi_0,\pi_1) = x_0 \rTo^{u_0} x_1 \ldots \in (X\times U)^\omega
\end{equation}
where for $k\ \ge\ 0$,
we have $u_k = \pi_0(x_0,\ldots,x_k)$, and $x_{k+1} = \pi_1(x_0,\ldots,x_k,u_k)$. 
Based on~(\ref{Outcome}) we define the infinite state behavior: 
$$states(x_0,\pi_0,\pi_1) = x_0 x_1 x_2\ldots \in X^\omega$$
and the corresponding outputs as: 
$$outputs(x_0,\pi_0,\pi_1) = H(x_0) H(x_1) H(x_2)\ldots\in Y^\omega.$$

For $i \in \{0,1\}$, given an initial state $x$ and a winning objective $\Phi \subseteq Y^\omega$,
we say the state $x \in X$ is \emph{winning} for player-$i$ if there is a player $i$
strategy $\pi_i$, such that, for all player-$(1 - i)$ strategies $\pi_{1-i}$, 
we have $outputs(x,\pi_0,\pi_1)\ \in\  \Phi$.  
The \emph{controller synthesis} problem asks, given a system $S$ and an
objective $\Phi\subseteq Y^\omega$, to construct a strategy $\pi$ for
player~0 such that every initial state $x_0$ is winning for $\Phi$, that
is, $outputs(x_0, \pi, \pi_1) \in \Phi$ for every $x_0\in X_0$ and
every player~1 strategy $\pi_1$.
In that case, $\pi$ is called a {\em controller} for $\Phi$, and
player~0 is said to {\em enforce} $\Phi$.

A {\em strategy-set} (for player~0) is a function $\hat{\pi}_0 : (X\times U)^* \times X \rightarrow 2^U$.
A strategy $\pi_0$ for player~0 is {\em compatible} with a strategy-set $\hat{\pi}_0$ if
for each $\mathsf{z} \in (X\times U)^*$ and $x\in X$, we have $\pi_0(\mathsf{z}\cdot x) \in \hat{\pi}_0(\mathsf{z}\cdot x)$.
A strategy-set $\hat{\pi}_0$ for player 0 is winning for a winning objective $\Phi$ if
every strategy compatible with $\hat{\pi}_0$ is winning for player 0.
A strategy-set $\hat{\pi}_0$ is {\em maximal} for $\Phi$ if it is winning for $\Phi$ and
every winning strategy of player~0 for $\Phi$ is compatible with $\hat{\pi}$.
A strategy-set $\hat{\pi}_0$ is memoryless if it only depends on the
final state and not the history of the play.
As with strategies, we represent a memoryless strategy-set as a
function $\hat{\pi}_0 : X \rightarrow 2^U$.

As an example, let $Z \subseteq Y$ and consider the property $\Phi$
to be the set of traces $Z^\omega$. 
This is called a {\em safety game}, and player~0 wins this game from $x$ if she
has a strategy $\pi_0$ such that for every strategy $\pi_1$ of player 1,
$outputs(x,\pi_0,\pi_1)$ is a trace consisting only of outputs in $Z$
(the game always remains in $Z$).
It is known that player~0 has a memoryless maximal strategy in a
safety game \cite{Zielonka98}.

For a set $X'\subseteq X$, define $\mathit{CPre}(X') = \set{x \in X
  \mid \exists u\in U. Post_u(x) \subseteq X'}$.
The set $\mathit{CPre}(X')$ consists of all states from which player~0
can force a visit to $X'$ in one step, no matter how player~1 resolves
the nondeterminism.
One can solve a safety game by iterating $\mathit{CPre}$, starting
from the set $H^{-1}(Z)$, until a fixpoint is reached \cite{MPS95,Zielonka98}:
\[
\nu x. H^{-1}(Z) \cap \mathit{CPre}(x)
\]
Indeed, this algorithm for solving safety games has been implemented
in several tools, including {\sc Pessoa}.

\subsection{Approximate Alternating Simulation}
\label{SSec:AB}

In order to capture the adversarial intent of the environment, the notion of equivalence and pre-order used in this paper is that of alternating simulation. Moreover, since 
the results in~\cite{PGT08,ZPT10} are used to relate differential equation models of physical systems to finite abstractions, we consider approximate alternating simulation relations.

\begin{definition}
\label{Def:ApproxAlternatingSimulation}
Let $S_a$ and $S_b$ be metric systems with $Y_a=Y_b$ and let $\varepsilon\in\R_0^+$. A relation \mbox{$R\subseteq X_a\times X_b$} is an \emph{$\varepsilon$}-\emph{approximate alternating simulation relation} from $S_a$ to $S_b$ if the following three conditions are satisfied:
\begin{enumerate}
\item for every $x_{a0}\in X_{a0}$ there exists $x_{b0}\in X_{b0}$ with $(x_{a0},x_{b0})\in R$;
\item for every $(x_a,x_b)\in R$ we have $\d(H_a(x_a),H_b(x_b))\le\varepsilon$;
\item for every $(x_a,x_b)\in R$ and for every $u_a\in U_a(x_a)$ there exists $u_b\in U_b(x_b)$ such that for every $x_b'\in \Post_{u_b}(x_b)$ there exists $x_a'\in\Post_{u_a}(x_a)$ satisfying $(x_a',x_b')\in R$.
\end{enumerate}
We say that $S_a$ is $\varepsilon$-approximately alternatingly simulated by $S_b$ or that $S_b$ $\varepsilon$-approximately alternatingly simulates $S_a$, denoted by $S_a \preceq_{\mathcal{AS}}^\varepsilon S_b$, if there exists an $\varepsilon$-approximate alternating simulation relation from $S_a$ to $S_b$. 
\end{definition}

The results in~\cite{PGT08,ZPT10} show that for any differential equation model of the physical world, 
it is possible to construct a finite system $S$ that is $\varepsilon$-approximate alternatingly simulated by the differential equation. Hence, once we synthesize a controller for the finite abstraction, such controller can be refined to a controller enforcing the same specification on the differential equation up to an error of $\varepsilon$. Note that $\varepsilon$ is a design parameter that can be made as small as desired, at the expense of a larger finite abstraction. 
In the remainder of the paper we will assume that we have already
abstracted the differential equation into a finite system. The
constructions of such abstractions has been implemented in the freely
available
tool {\sc Pessoa} \cite{MazoDT10}.

\section{Specifications}

\subsection{Linear Temporal Logic}

We now review the syntax and semantics of linear-temporal logic (LTL) \cite{Pnueli77}.

\begin{definition}
The set of LTL formulae is generated by the following grammar:
\[
\varphi \, ::=\, p \mid \varphi \vee \varphi \mid \varphi\wedge \varphi \mid \lnot \varphi \mid \bigcirc\; \varphi \mid \varphi\;\U\; \varphi \mid \varphi\; \W\; \varphi
\]
where $p$ is chosen from a set $\mathcal{P}$ of atomic propositions.
\end{definition}
We define shorthands 
$\mathbf{true}$ and $\mathbf{false}$ as shorthand for $p \vee \lnot p$ and $p\wedge \lnot p$ respectively.
We use $\Diamond \varphi$ and $\Box \varphi$  as shorthands of $(\mathbf{true}\ \U\ \varphi)$ and $(\varphi\ \W\ \mathbf{false})$ respectively.

An LTL formula is in {\em negation normal form (NNF)} if negation occurs only before the atomic
propositions.
It is known that any formula can be put in NNF by applying de Morgan's laws (for Boolean operations),
and the identities 
$\lnot\lnot \varphi \equiv \varphi$,
$\lnot \bigcirc \varphi \equiv \bigcirc \lnot \varphi$, and 
$\lnot (\varphi_1\W\varphi_2) \equiv \lnot \varphi_2 \U \lnot \varphi_2 \wedge \lnot \varphi_1$.
The length $|\varphi|$ of a formula $\varphi$ is the number of symbols in $\varphi$ and defined by induction
on the structure of $\varphi$ in a standard way.

The semantics of LTL formulae is defined 
over infinite sequences $\mathsf{z}\in (2^{\mathcal{P}})^\omega$:
\begin{itemize}
\item $\mathsf{z}\models p$ iff $p\in \mathsf{z}(0))$;
\item $\mathsf{z}\models \lnot\varphi$ iff $\mathsf{z} \not\models \varphi$;
\item $\mathsf{z}\models \varphi{\land}\psi$ iff $\mathsf{z}\models \varphi$ and $\mathsf{z}\models\psi$;
\item $\mathsf{z}\models \varphi\lor\psi$ iff $\mathsf{z}\models \varphi$ or $\mathsf{z}\models\psi$;
\item $\mathsf{z}\models \bigcirc\,\varphi$ iff $\mathsf{z}[1]\models \varphi$;
\item $\mathsf{z}\models \varphi\,\mathsf{U}\,\psi$ iff $\exists k\ge 0$ s.t. $\mathsf{z}[k]\models \psi$ and $\mathsf{z}[j]\models \varphi$ for all $0\le j<k$.
\item $\mathsf{z}\models \varphi\,\mathsf{W}\,\psi$ iff $\mathsf{z}[i]\models \varphi$ for all $i\in\mathbb{N}_0$ or $\exists k\ge 0$ $\mathsf{z}[k]\models \psi$ and $\mathsf{z}[j]\models \varphi$ 
for all $0\le j<k$.
\end{itemize}
If $\mathsf{z}\models\varphi$, we say $\mathsf{z}$ \emph{satisfies} $\varphi$.
For an LTL formula $\varphi$, the language $L (\varphi)$ of all strings satisfying $\varphi$ is defined by: 
$$L(\varphi) = \{ \mathsf{z} \in (2^{\mathcal{P}})^{\omega} \mid \mathsf{z} \models \varphi\}.$$ 

Let $S$ be a system where $Y = 2^{\mathcal{P}}$ and thus $H$ maps each state $x\in X$ to the set of atomic propositions that are true at $x$.
We say player~0 enforces the LTL formula $\varphi$ if 
there exists a player~0 strategy $\pi_0$ such that for each player~1 strategy $\pi_1$ and each $x_0\in X_0$
we have that  $outputs(x_0,\pi_0,\pi_1)$ satisfies $\varphi$.

%% file: safe-ltl.tex
\subsection{Safe-LTL}

We now define a subset of LTL formulas that capture all safety properties.

\begin{definition}
The set of safe-LTL formulae is generated by the following grammar:
\[
\varphi\, ::= \, p \mid \lnot p \mid \varphi \vee \varphi \mid \varphi \wedge \varphi \mid \bigcirc\; \varphi \mid \varphi\; \W\; \varphi
\]
where $p$ ranges over a set $\mathcal{P}$ of atomic propositions.
\end{definition}

\begin{figure*}[htb]
\centering
\subfigure[Fine Automaton for $p\ \W\ q$.]{
\includegraphics[scale=0.6]{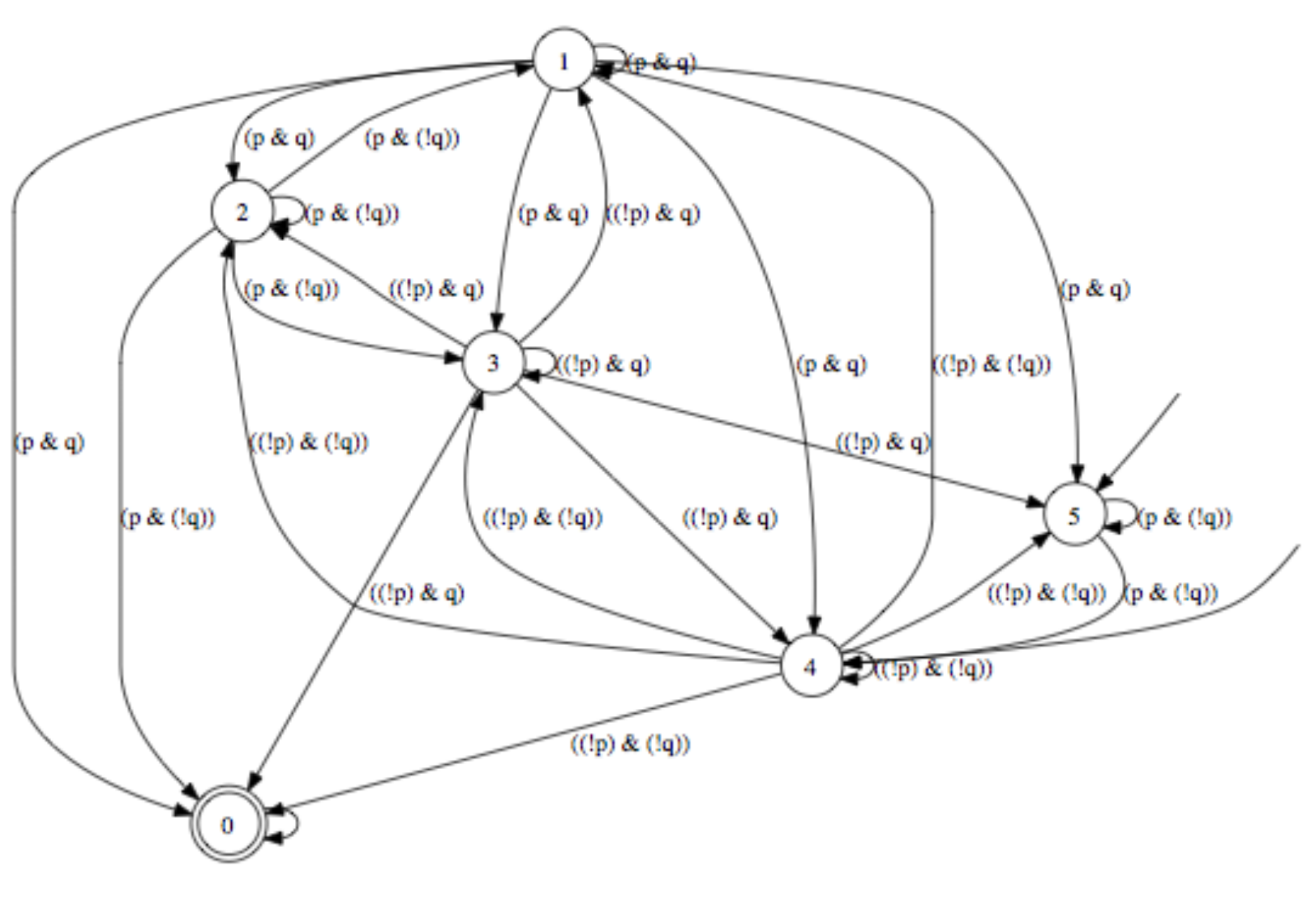}
\label{fig:pwqnondet}
}
\subfigure[Determinized Version]{
\includegraphics[scale=0.5]{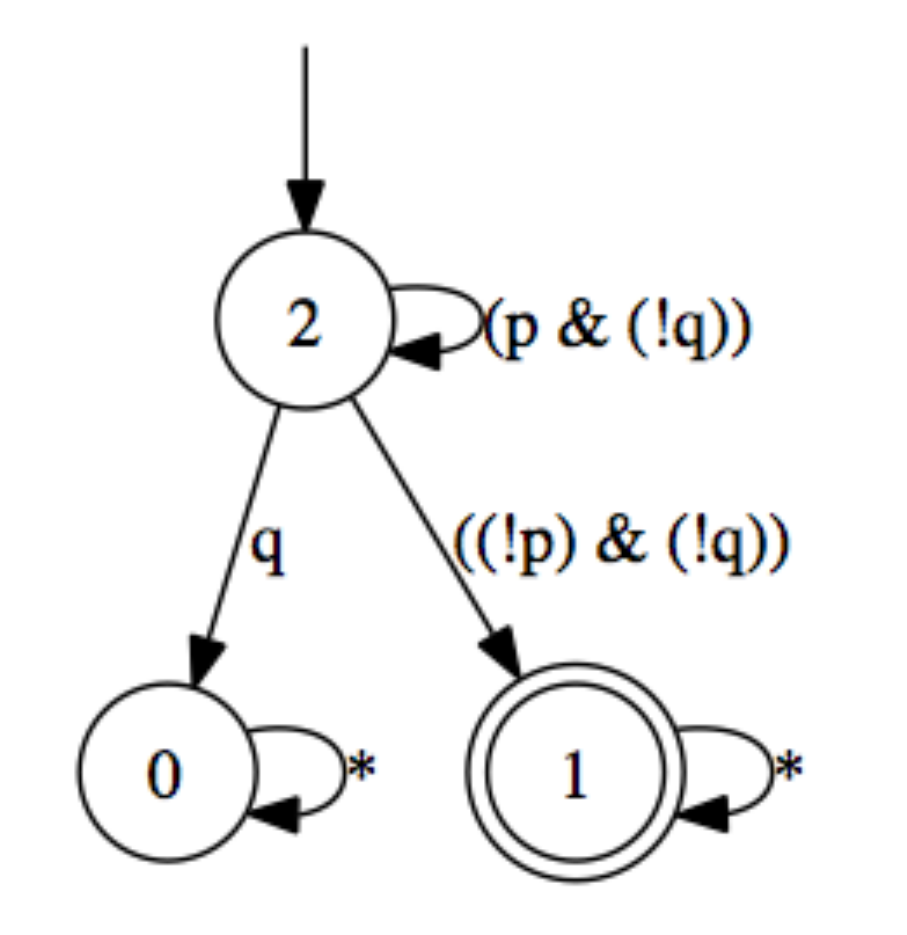}
\label{fig:pwqdet}
}
\label{fig:pwq}
\caption[optional]{$p\ \W\ q$}
\end{figure*}

A safe-LTL formula always defines a safety property. 
Intuitively, a formula $\varphi$ defines a safety property if $\mathsf{z}\not\models \varphi$ can 
be checked by looking at a finite prefix of $\mathsf{z}$.

Thus, reasoning about safety properties on infinite behaviors can be reduced to reasoning
about their finite prefixes.
First, we recall nondeterministic finite automata as acceptors of languages over finite words.
A \emph{nondeterministic finite automaton (NFA)} is a 5-tuple 
$A$ = $(Q$, $Q_0$, $\Sigma$, $\delta$, $F)$,
where $Q$ is a finite set of states, $Q_0 \subseteq Q$ is a set of initial states, $F \subseteq Q$
is a set of final states, $\Sigma$ is an alphabet, and $\delta \subseteq Q \times \Sigma \times Q$ is a set of transitions.
An NFA is \emph{deterministic}, written DFA, if $|Q_0|\ =\ 1$ and $\delta$ defines a total function
from $Q \times \Sigma$ into $Q$. 
The unique successor of a state $q\in Q$ under the letter $\sigma\in \Sigma$ 
in a deterministic automaton is denoted by $\delta (q,\sigma)$. A {\em run} of an NFA on a word $\sigma \equiv \sigma_0\ldots\sigma_{n-1}\in\Sigma^*$
is a sequence $q_0 \xrightarrow{\sigma_0} q_1 \ldots q_{n-1} \xrightarrow{\sigma_{n-1}} q_n$ such that $q_0\in Q_0$ and
for each $0\leq i\leq n-1$ we have $(q_i,\sigma_i, q_{i+1})\in\delta$.
A run is accepting if moreover $q_n \in F$, and we say the NFA accepts $\sigma$.
The language of an NFA is the set of all words $\sigma\in \Sigma^*$ such that the NFA has an accepting run on $\sigma$.

The set of \emph{bad prefixes} for a safety formula $\varphi$ is defined by:
$$Bad(\varphi) = \{\mathsf{z} \in (2^{\mathcal{P}})^{*} \mid \forall \mathsf{w} \in (2^{\mathcal{P}})^{\omega}\,\,  \mathsf{z}.\mathsf{w} \not\models \varphi\}.$$
That is, a (finite) prefix $\mathsf{z}$ is bad if none of its infinite extensions $\mathsf{z}\cdot\mathsf{w}$ 
satisfies the formula $\varphi$.
The set of fine prefixes is the set of finite prefixes that are sufficient to prove that the computation is unsafe.
We say that a set $Z \subseteq Bad(\varphi)$ is a \emph{trap} for the safety language $L(\varphi)$ iff
every word $\w \notin L(\varphi)$ has at least one prefix $\z \in Z$.  
We denote all the traps for $L(\varphi)$ by $trap(L(\varphi))$.

We say that a nondeterministic automaton $N_{\psi}$ is \emph{fine} for $\psi$ iff 
there exists $Z \in trap(L(\psi))$ such that $L(N_{\psi}) = Z$.
Thus, a fine automaton $N_{\psi}$ may not accept all the bad prefixes, however
it should accept at least one bad prefix of every computation that does not satisfy $\psi$.
%

Kupferman and Vardi \cite{KV01,KupfermanL06} show that an automaton fine for $\varphi$ can
be constructed from $\varphi$.
The translation is based on the \emph{reverse deterministic} automaton defined 
in~\cite{VardiW94}. 
In \tool~we implemented the version of Kupferman and Vardi's algorithm reported in~\cite{Lat03} and 
presented here as Algorithm~\ref{algo:fine}. 
This algorithm computes $N_{\varphi}$ from a safe-LTL formula $\varphi$. 
It first computes the set of subformulas $cl$ of $\lnot \varphi$ by the procedure
\emph{computeClosure}.
Since each state of the automaton represent whether each of the subformulas is either 
true or false in that state, the fine automaton can have at most $2^{|cl|}$ states.

\begin{proposition}
For every safe-LTL formula $\varphi$, Algorithm~\ref{algo:fine} constructs a
nondeterministic fine automaton for $\varphi$ with at most $2^{|\varphi|}$ states. 
\end{proposition}

\begin{algorithm}[t]
\begin{algorithmic}
\begin{tabbing}
\= $\psi'\ :=\ NNF(\neg\psi)$; $cl\ :=\ computeClosure(\psi')$\\
\> $F := \{\emptyset\}$, $Q := \{\emptyset\}$; $X := \{\emptyset\}$, $Q_0 := \{\}, \delta = \{\}$\\ 
\> {\bf while} $X \neq \emptyset$ {\bf do}\\
\> \quad $s := Dequeue(X)$\\
\> \quad {\bf foreach} $\sigma \in \Sigma$\\
\> \quad \quad $s' = \{\}$\\
\> \quad \quad {\bf foreach} $\phi \in cl$ {\bf do}\\
\> \quad \quad \quad {\bf switch} $\phi$ {\bf begin}\\
\> \quad \quad \quad \quad {\bf case} $p = q$ or $p = \lnot q$ for $q \in Y$:\\
\> \quad \quad \quad \quad if $p$ is satisfied by $\sigma$, then $s' := s' \cup \{p\}$\\
\> \quad \quad \quad \quad {\bf case} $\phi = \phi_1 \lor \phi_2$ :\\
\> \quad \quad \quad \quad if $\phi_1 \in s'$ or $\phi_2 \in s'$ then $s' := s' \cup \{\phi\}$\\
\> \quad \quad \quad \quad {\bf case} $\phi = \phi_1 \land \phi_2$ :\\
\> \quad \quad \quad \quad if $\phi_1 \in s'$ and $\phi_2 \in s'$ then $s' := s' \cup \{\phi\}$\\
\> \quad \quad \quad \quad {\bf case} $\phi = \bigcirc \phi_1$ :\\
\> \quad \quad \quad \quad if $\phi_1 \in s$ then $s' := s' \cup \{\phi\}$\\
\> \quad \quad \quad \quad {\bf case} $\phi = \phi_1 \U \phi_2$ :\\
\> \quad \quad \quad \quad if $\phi_2 \in s'$ or ($\phi_1 \in s'$ and $\phi \in s$)\\
\> \quad \quad \quad \quad then $s' := s' \cup \{\phi\}$\\
\> \quad \quad \quad {\bf end switch}\\
\> \quad \quad {\bf end for}\\
\> \quad \quad {\bf if} $\lnot\varphi \in s'$ {\bf then} $Q_0 := Q_0 \cup \{s'\}$\\
\> \quad \quad $\delta := \delta \cup \{(s', \sigma, s)\}$\\
\> \quad \quad $X := X \cup \{s'\}$, $Q := Q \cup \{s'\}$\\
\> \quad {\bf end for}\\
\> {\bf end while}\\
\> {\bf return} $A^{Fine}_{\lnot \varphi} = (Q,Q_0,2^{\mathcal{P}},\delta,F)$
\end{tabbing}
\end{algorithmic}
\caption{
{\bf ConstructFineAutomaton($\psi$)}\label{algo:fine}}
\end{algorithm}

\section{Controller Synthesis}

In this section, we assume that we have already computed a finite
abstraction, 
in the form of a system $S$, of the physical components.
\tool~accepts a pair of specifications $(\varphi_S,\varphi_L)$: the
first, $\varphi_S$, is a safe-LTL
formula that specifies the safety requirements of the system, and
the second, $\varphi_L$, is a guarantee formula of the form $\Diamond p$ that
specifies that the goal $p$ is eventually reached.
We perform controller synthesis in two steps.
First, we compute the maximal winning strategy for player~0 for the
safe-LTL part of the specification.
Second, we compute a controller that ensures the guarantee property
using a strategy compatible with the maximal strategy.

\subsection{Controller Synthesis for Safe-LTL}

For synthesizing a controller for a safe-LTL formula $\varphi$, we construct a
deterministic automaton on finite words that is fine for $\varphi$.
Note that Algorithm~\ref{algo:fine} may produce an NFA.
However, determinization for NFAs over finite words uses the (easier
to implement) subset construction.


Theoretically, the determinization step adds one more exponential,
making the complexity of the construction doubly exponential in the
size of $\varphi$.
In our practical examples, this double exponential behavior has not
shown up.
For example, given the fine automaton for $p\, \mathsf{W}\, q$, the subset construction
creates the deterministic automaton Figure~\ref{fig:pwqdet}.

Given a system $S = (X, X_0, U, \rTo, Y, H)$ and a DFA $D_{\varphi} =
(Q, q_0, Y, \delta, F)$ fine for $\varphi$, we
define the \emph{synchronous product} $S\times D_{\varphi} =
(X',X'_0,U',\rTo', Y', H')$ where
\begin{itemize}
 \item $X' = X \times Q$;
 \item $X'_0 = \{(x,q) \mid x \in X_0, q = \delta(q_0,H(x))\} $;
\item  $U' = U$;
\item  $(x,q)\to'^u (x',q')$ if $x \to^u x' $ and $\delta(q,H(x')) = q'$;
\item $Y' = Y$;
\item  $H'((x,q)) = H(x)$ for each $(x,q) \in X'$.
\end{itemize}
A controller enforcing $\varphi$ on $S$ can be constructed by synthesizing a controller
on the synchronus product $S\times D_{\varphi}$ enforcing the
specification that the system always remains in the states 
$X \times (Q\setminus F)$, i.e., that player~0 ensures that no word in the
language of $D_{\varphi}$ is seen.
This is a safety game where player~0 keeps the states into an
invariant set ($X\times (Q\setminus F)$), and can be solved using
existing methods by iterating a symbolic controllable-predecessor operator
\cite{Zielonka98,MazoDT10}.
Moreover, it is well-known that player~0 has memoryless maximal
winning strategies in this game.

%
%
%

\begin{theorem}\label{th:fine}
Let $S =  (X,X_0,U,\rTo,Y,$ $H)$ be system and let $D_{\varphi} = (Q,Q_0,Y, \delta,F)$ 
be a deterministic finite automaton fine for the safe-LTL formula $\varphi$.
For any initial state $x\in X_0$, player 0
has a winning strategy for the safe-LTL formula $\varphi$, if
player 0 has a memoryless winning strategy from the unique $x_0 \in X'_0$ to stay in $X
\times (Q\setminus F)$ states in system $S\times D_{\varphi}$.
Moreover, player~0 has a maximal winning strategy in $S\times D_\varphi$.
\end{theorem}

Thus, the algorithm to construct a maximal memoryless controller for a
system $S$ and a 
safe-LTL property $\varphi$ proceeds as follows.
First, we construct an NFA $N_\varphi$ fine for $\varphi$.
Second, we use the subset construction to determinize $N_\varphi$ into
a DFA $D_\varphi$.
Third, we construct the synchronous product of $S$ with $D_\varphi$.
Finally, we solve the safety game on $S\times D_\varphi$ for the
winning set $X\times (Q\setminus F)$ and construct a maximal
memoryless winning strategy.

\subsection{Controller Synthesis for the Guarantee Part}

Let $S\times D_\varphi = (X, X_0, U, \rTo, Y, H) $ be the synchronous product of a system and a
DFA fine for the safe-LTL $\varphi$, and let $\pi$ be a maximal
memoryless winning strategy for player~0 which ensures that all runs of the
system stay in the states $X\times (Q\setminus F)$.

We define the restriction of $S\times D_\varphi$ modulo $\pi$ to be
the system $(X,X_0, U, \rTo', Y, H)$ where $x\rTo'^{u} x'$ if
$x\rTo^{u} x'$ and $u \in \pi(x)$.
That is, we restrict the actions available at a state to only those
allowed by the maximal strategy $\pi$.

We now consider constructing a controller for the guarantee part
$\Diamond p$.
We solve this by constructing a winning strategy in the reachability
game on the product $S\times D_\varphi$ modulo $\pi$, the maximal
memoryless winning strategy for the safety game.
Again, the solution to the reachability game is constructed by
iterating a symbolic controllable predecessor operator
\cite{Zielonka98,MazoDT10}.

The resulting strategy ensures that the guarantee part $\Diamond p$ is
enforced by player~0 (by construction in the reachability game), 
while always maintaining the safety part (by ensuring that the
strategy is compatible with $\pi$).
Together, the controller enforces the specification $\varphi_S \wedge \varphi_L$.

While the current implementation of \tool~only handles guarantee
properties of the form $\Diamond p$ (or some syntactic sugar, e.g.,
properties of the form $p_1 \U p_2$ using the identity $p_1 \U p_2
\equiv p_1\W p_2 \wedge \Diamond p_2$), notice that all we need is
that a deterministic generator for the liveness part of the
specification is efficiently computable.
For example, it is easy to extend the algorithm when the liveness part
of the specification is a B\"uchi requirements $\Box\Diamond p$, or
more generally, from the fragments described in \cite{AlurT04}.

%% file: refinement.tex
\section{Controller Refinement}
The discussion so far has focused on the synthesis of strategies enforcing LTL formulas over the finite abstraction $S$ of a physical system. The natural next step is to refine the controller synthesized for $S$ to a controller enforcing the specification on the differential equation model of the physical system. Typical controller implementations are done on digital platforms, hence it is convenient to assume a periodic\footnote{There are also considerable advantages to consider non-periodic implementations as in~\cite{AT09}, however such approaches are outside the scope of this paper.} execution of the controller implementation with period $\tau$. Moreover, a time discretized version of the differential equation:
\begin{equation}
\label{DiffEq}
\dot{x}=f(x,u),\qquad x\in \R^n,\quad u\in\R^m
\end{equation}
modeling the physical system being controlled can be described by the system $S_\tau=(X_\tau,X_{\tau 0},U_\tau,\rTo_\tau,Y_\tau,H_\tau)$ consisting of:
\begin{itemize}
\item $X_\tau=\R^n$;
\item $X_{\tau 0}=X$;
\item $U_\tau=\R^m$;
\item $x\rTo^u_\tau x'$ if there exists a solution $\xi$ of~(\ref{DiffEq}) for the constant input $u$ satisfying $\xi(0)=x$ and $\xi(\tau)=x'$.
\item $Y_\tau=X_\tau$;
\item $H_\tau(x)=x$ for any $x\in X_\tau$.
\end{itemize}

The results in~\cite{PGT08,ZPT10} guarantee the existence of a finite system $S$ and of an $\varepsilon$-approximate alternating simulation relation $R$ from $S$ to $S_\tau$. Note that while $S_\tau$ is deterministic, the abstraction process introduces nondeterminism in $S$. Nevertheless, the existence of the relation $R$ guarantees that any controller synthesized for $S$ can be refined to a controller for $S_\tau$. A formal description of the refined controller can be found in~\cite{Tab09}. Here, we provide an informal description which we believe to be more  informative. Any state $x_\tau\in X_\tau$ of the system $S_\tau$ is related by $R$ to a state $x\in X$ in the finite abstraction $S$. If the strategy $\pi_0$ dictates that the input $u\in U$ should be used at the state $x$, then by using a constant input curve of duration $\tau$ and value $u$ in $S_\tau$, we are guaranteed to reach a state $x_\tau'\in X_\tau$ that is $R$ related to a state $x'\in \Post_u(x)$. Hence, the refined controller consists in a loop performing the following steps:
\begin{enumerate}
\item Acquire the current state from sensors/estimators;
\item Identify the state in $S$ that is related by $R$ to the current state;
\item Compute the input $u$ given by the strategy $\pi_0$;
\item Send the value $u$ to the actuators and keep it constant for $\tau$ units of time;
\item Loop to step 1.
\end{enumerate}

This refined controller enforces the specification on $S_\tau$ up to an error $\varepsilon$ as stated in the next result.

\begin{proposition}
Let $S_\tau$ be the time discretization of a differential equation governing the physical system to be controlled and let $\varphi$ be a LTL formula whose predicates correspond to subsets of $Y_\tau$. Consider the finite abstraction $S$ of $S_\tau$ and let $R$ be the $\varepsilon$-approximate alternating simulation relation from $S$ to $S_\tau$. For any strategy $\pi_0$ enforcing $\varphi$ on $S$, the strategy $\pi_{\tau 0}$ obtained by refining $\pi_0$, enforces $\varphi$ on $S_\tau$ up to an error of $\varepsilon$, that is, for any environment strategy $\pi_1$ for $S$ we have $\d(\mathsf{y}(i),\mathsf{y}_\tau(i))\le\varepsilon$ for every $i\in \N$, for the unique $\mathsf{y}\in outputs(x,\pi_0,\pi_1)$, the unique $\mathsf{y}_\tau\in outputs(x_\tau,\pi_{\tau 0})$, and for any $(x,x_\tau)\in R$.
\end{proposition}

%% file: casestudy.tex
\section{Case Study : Robot Controller}
 We consider a nonholonomic robot described by the following differential equations:
$$\dot{x}  =  v\cos\theta,\ \ \dot{y}  =  v\sin\theta,\ \ \dot{\theta}  =  \omega$$
\begin{figure}[htb]
\includegraphics[scale=0.51]{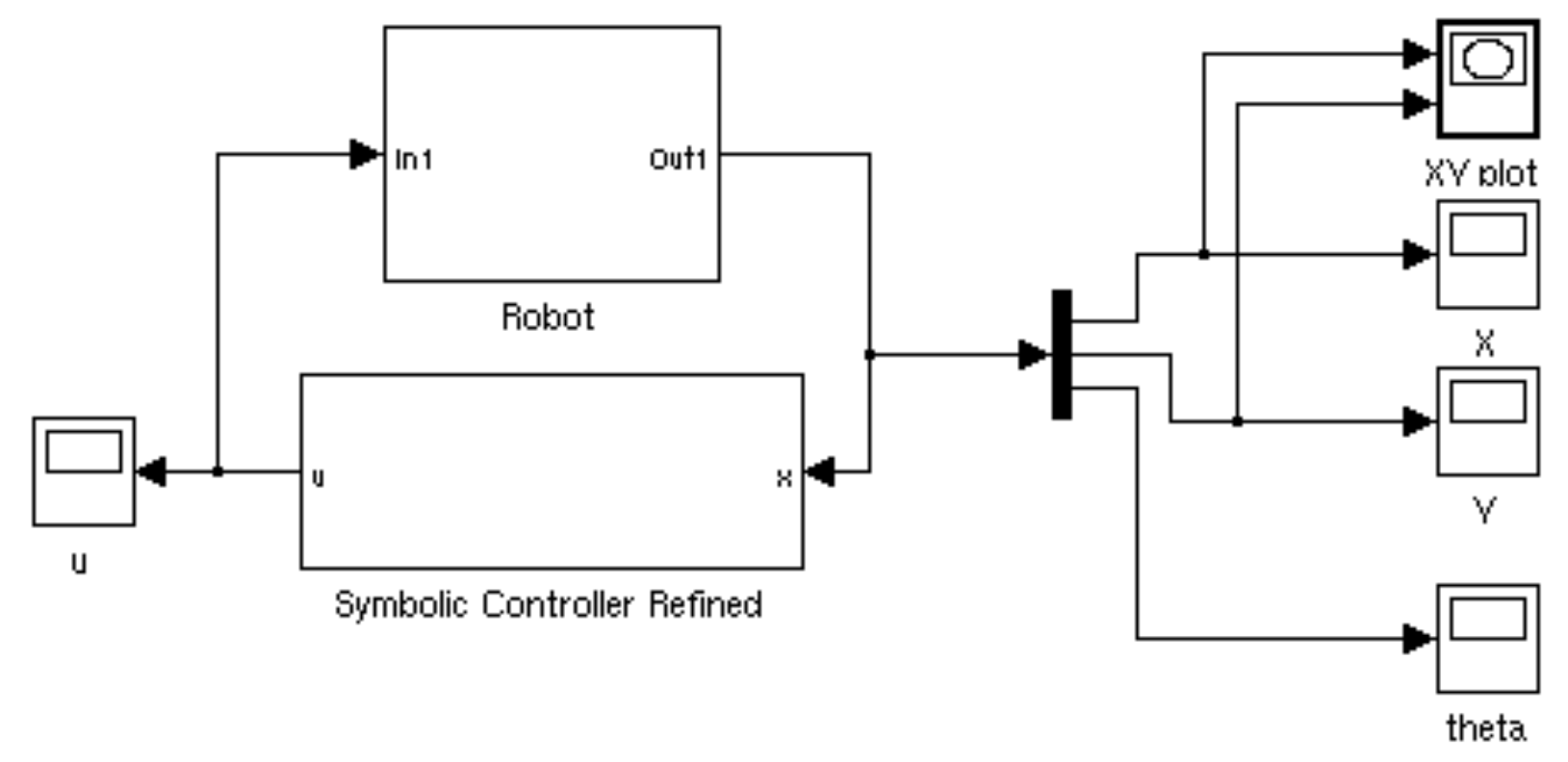}
\caption{Closed-loop diagram in Simulink showing the automatically synthesized controller.}\label{fig:model}
\end{figure}
where $(x,y)$ denotes the robot position and $\theta$ its orientation. The inputs are $v$ and $\omega$ and correspond to the linear and angular velocity of the robot, respectively. Using {\sc pessoa} we compute a finite abstraction $S$ of the differential equation model of the robot. This abstraction is approximately alternatingly simulated by the differential equation model with a precision of $\varepsilon=0.1$. In this abstraction the input $v$ is restricted to take values in the set $\{0,0.2,0.4\}$ while the input $\omega$ is restricted to take values in the set $\{-0.2,0,0.2\}$. 

\subsection{Reachability with Obstacle Avoidance}
\begin{figure}[htb]
\centering
\includegraphics[scale=0.5]{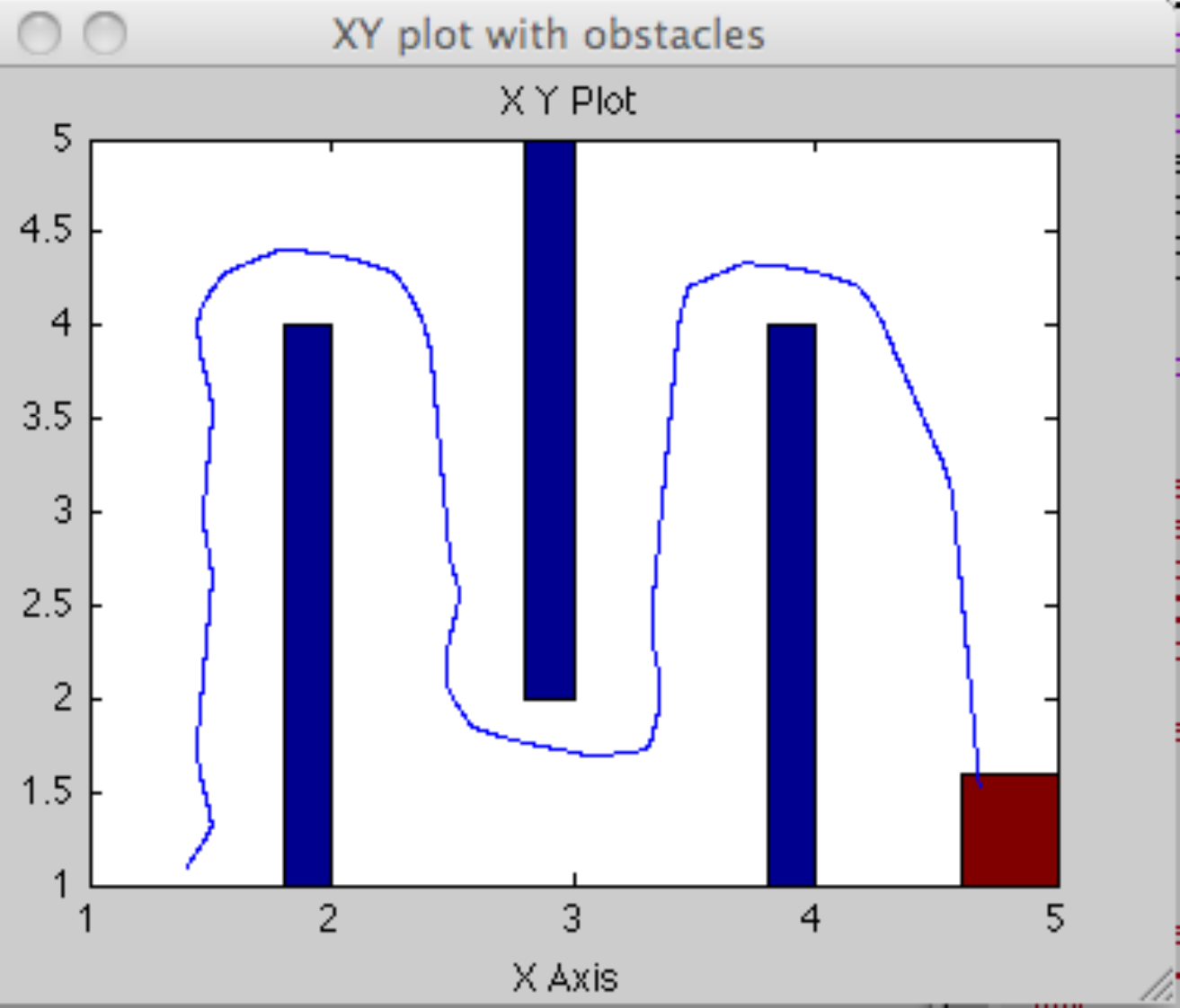}
\caption{Trajectory followed by the robot.}\label{fig:traj}
\end{figure}

\begin{figure}[htb]
\centering
\includegraphics[scale=0.35]{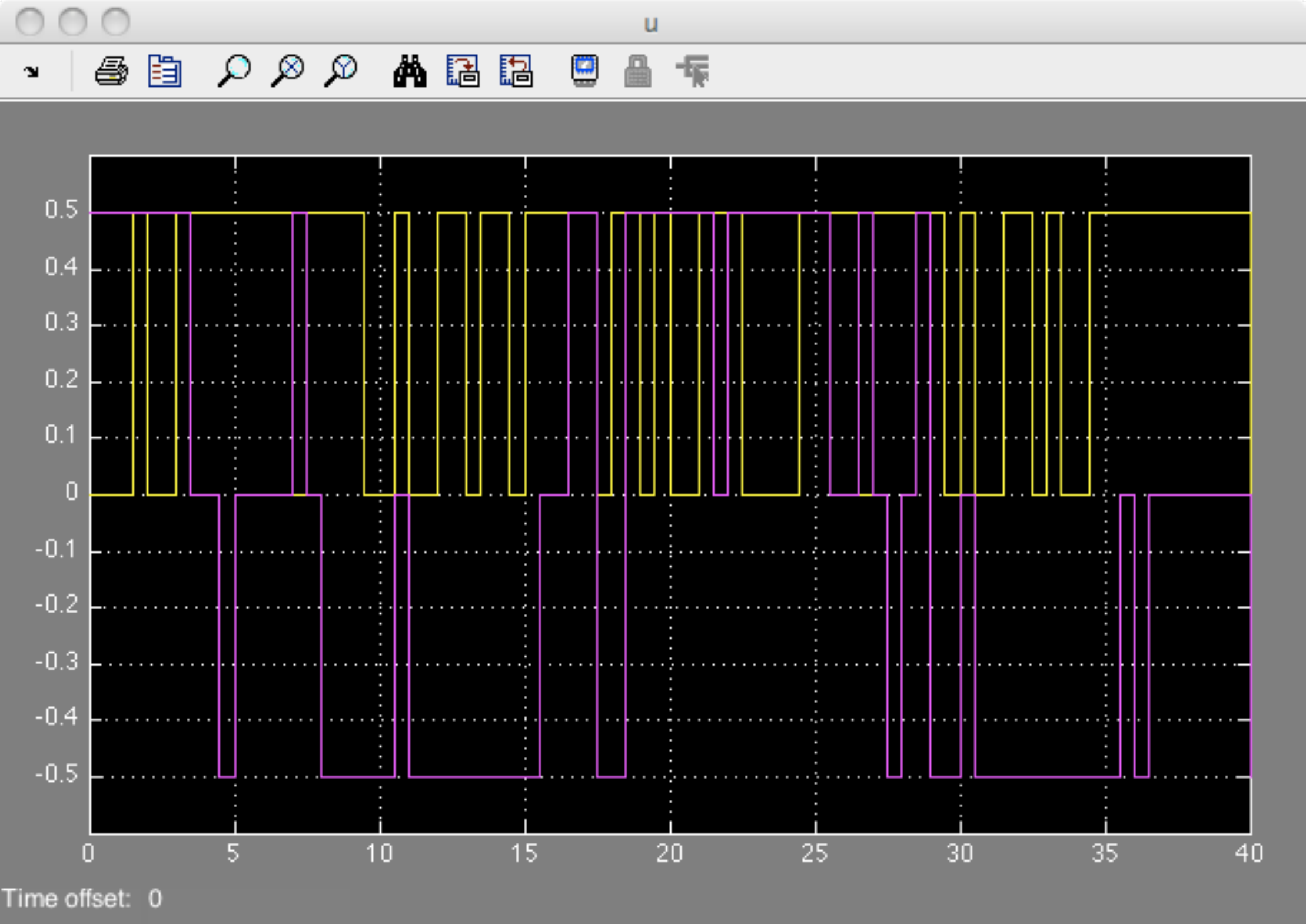}
\caption{Input signal generated by the controller.}\label{fig:u}
\end{figure}

For every obstacle (see the blue sets in Figure~\ref{fig:traj}) we construct a predicate $obstacle_i$, $i\in \{1,2,3\}$, that is true whenever the robot is inside the set defined by the obstacle. Similarly, we defined the predicate $target$ describing the target set represented by the red set in Figure~\ref{fig:traj}. The objective of reaching the target set, if possible, while avoiding the obstacles is naturally expressed by the safe-LTL formula:
$$\psi=(\neg (obstacle_1\lor obstacle_2\lor obstacle_3))\,\mathsf{W}\,target.$$
Note that $\varphi$ does not require the target set to be reached. Such requirement can be prescribed by using instead the LTL formula:
$$\varphi=(\neg (obstacle_1\lor obstacle_2\lor obstacle_3))\,\mathsf{U}\,target.$$
Since $\varphi$ can be decomposed as:
$$\varphi=\psi\,\land\,\Diamond \,target$$
we first solve the safety problem specified by $\psi$ and then we solve the reachability problem specified by $\Diamond \,target$. The synthesized controller is automatically refined to a Simulink block in {\sc pessoa}, see Figure~\ref{fig:model}, in order to simulate the closed-loop behavior. In Figure~\ref{fig:traj} we show the trajectory followed by the robot, and in Figure~\ref{fig:u} we show the inputs used to steer the robot. The yellow line represents the translational velocity input while the magenta line represents the angular velocity input.


\subsection{Fault tolerance}
We consider the same robot as in the previous case study. We assume that the communication between the several sensor onboard of the robot with the microprocessor running the control code is governed by a protocol that reports if communication is successful or not. There are several reasons for unsuccessful communication such as the fact that the communication medium is shared among several subsystems and sensor failures. We now consider a specification detailing how the robot should operate in case of sensor failures.
\begin{figure}[htb]
\vspace{-0.5cm}
\centering
\includegraphics[scale=0.35]{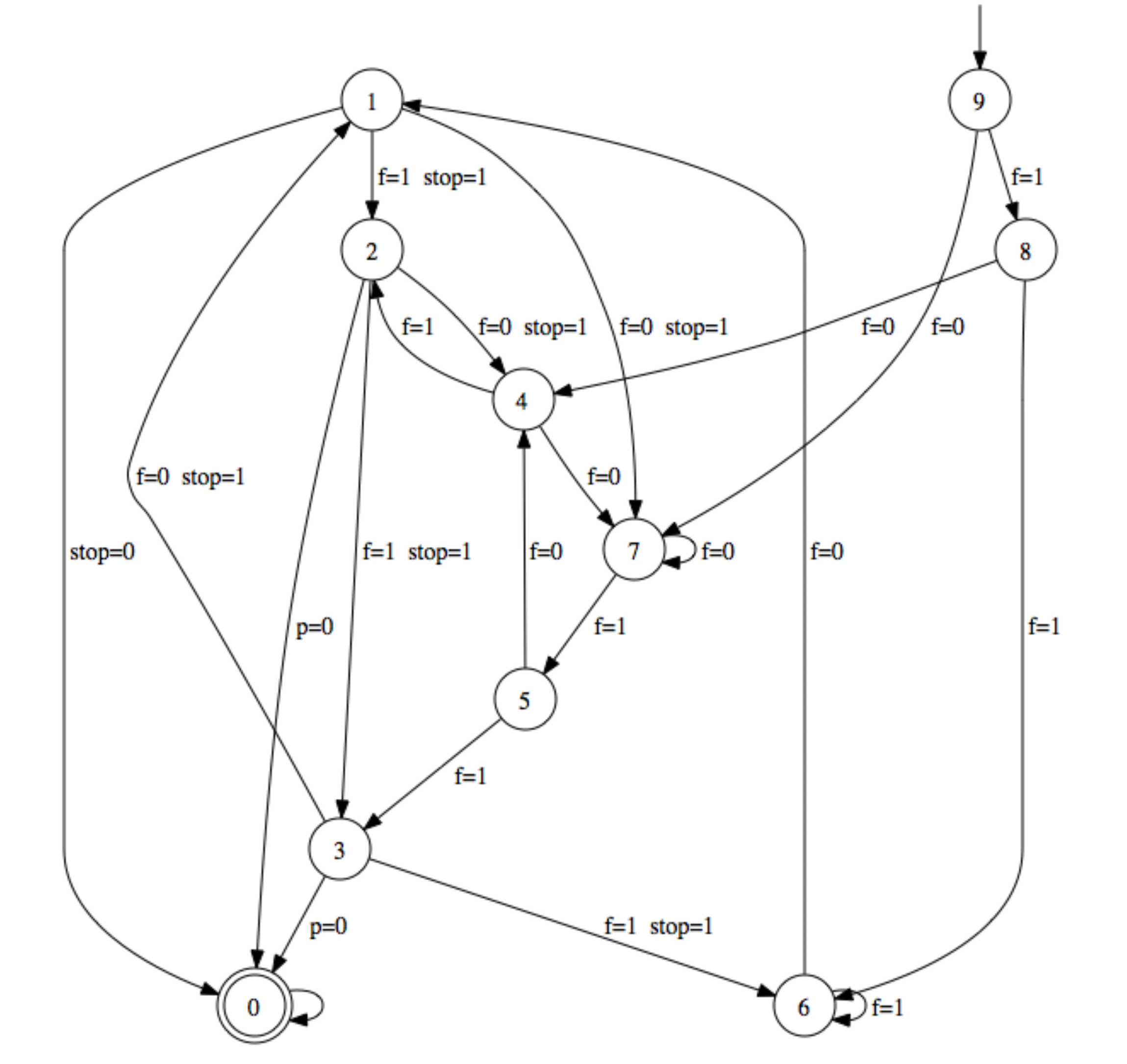}
\caption{Fine-automaton for the LTL formula~(\ref{LTLFault})}\label{fig:faultprefix}
\end{figure}

The main microprocessor may fail to receive sensor measurements more than once. In such case the controller should have a strategy to protect the robot from either leaving the desired working area or hitting the obstacles. One possible way of encoding this objective as a safety property is to require that if sensor measurements are not received two or more times during three consecutive control cycles, the robot should stop and remain at its current location. In order to formalize this property we extend the model of the robot so as to incorporate the previously used input as part of the state. Consider now the predicate $stop$, which is true (resp. false) when the input $v$ is equal to (resp. different from) zero, and the predicate $fail_{3,2}$, which is true when 2 or more sensor measurements were not received during 3 consecutive control cycles. Since in LTL we cannot refer to the past, we encode $fail_{3,2}$ by making reference to the future as follows:
$$fail_{3,2} = (f \land \bigcirc f) \lor (\bigcirc f \land \bigcirc \bigcirc f) \lor (f \land \bigcirc\bigcirc f).$$

\begin{figure}[htb]
\centering
\includegraphics[scale=0.37]{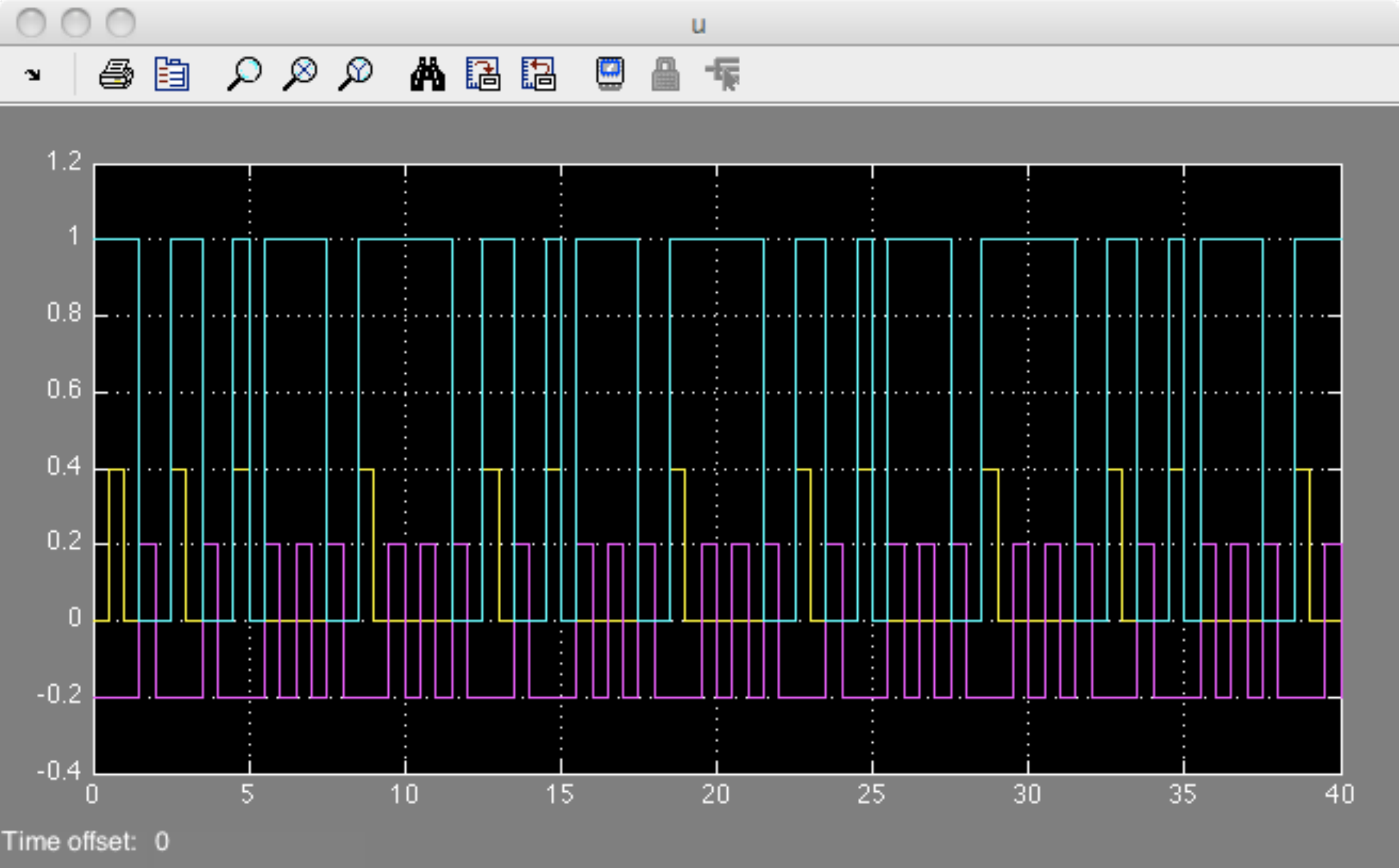}
\caption{Inputs Generated by Controller}\label{fig:fu}
\end{figure}

\begin{figure}[htb]
\centering
\includegraphics[scale=0.37]{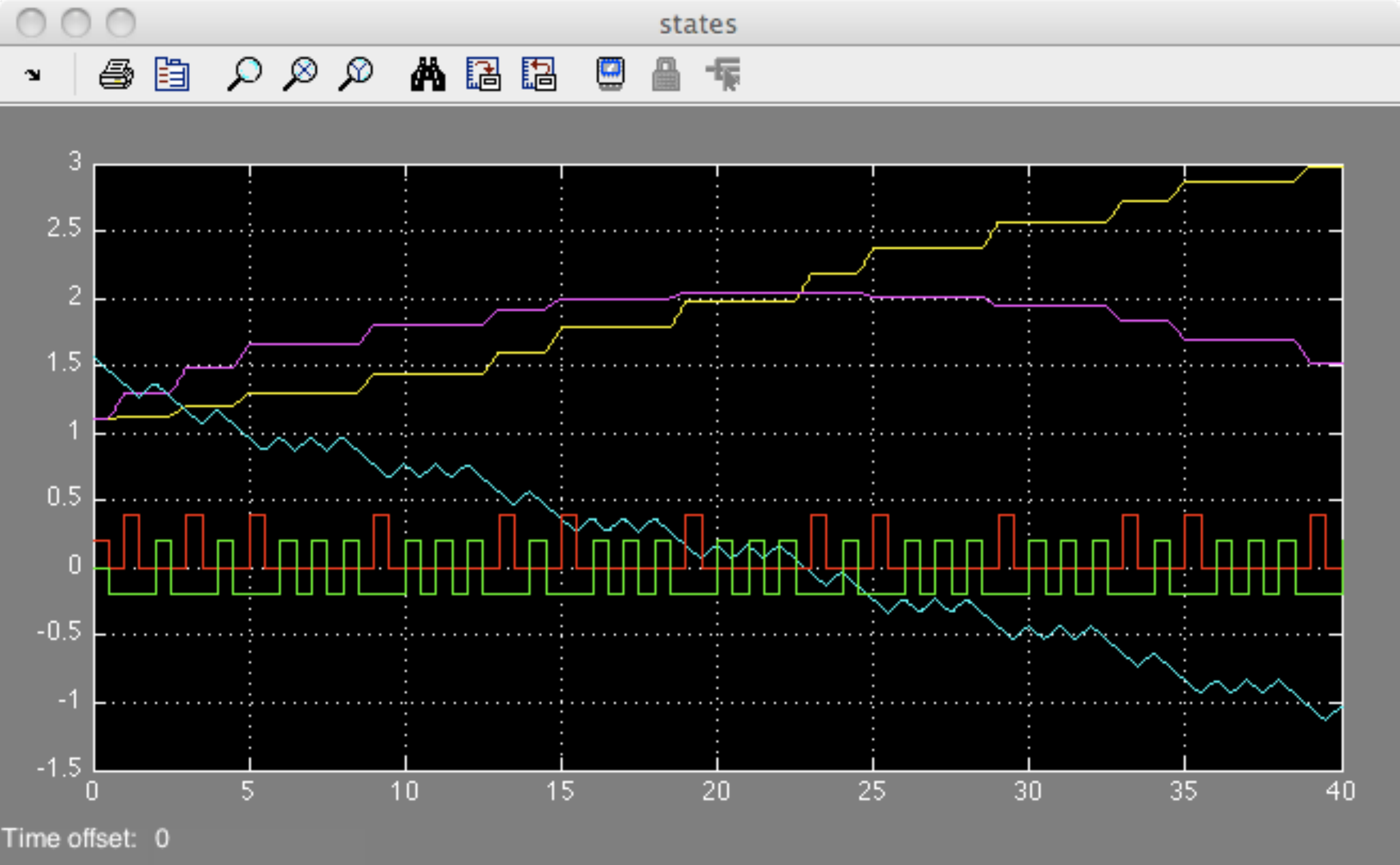}
\caption{States}\label{fig:states}
\end{figure}

In the preceding formula $f$ is the predicate that becomes true every time that the microprocessor fails to receive sensor measurements. The final formula can then be obtained as:
\begin{equation}
\label{LTLFault}
\Box (fail_{3,2} \to \bigcirc\bigcirc\bigcirc\, stop).
\end{equation}
Figure~\ref{fig:faultprefix} shows the fine-automaton with respect to the previous property. In 
Figure~\ref{fig:fu} we show the inputs generated by the controller when the predicate $f$ evolves according to:
$$f\,f\,f\,\neg\,f\,\neg f\,f\,f\,\neg f\,\neg f\,f\,\neg f\,f\,f\,f.$$
 The yellow line represents the translational velocity input ($v$) while the magenta line represents the angular velocity input ($\omega$).
Note that whenever the protocol returns two consecutive failures ($f$ is true twice), the input $v$ generated by the controller at the next control cycle is zero.
Figure~\ref{fig:states} shows the closed-loop evolution of $\theta$, $x$, $y$, $u$ and $v$ for the given fault-sequence.
The colors of these state variables are cyan, yellow, magenta, red and green respectively.
We can easily develop more sophisticated fault tolerance requirements. Let $slow$ denote the predicate that holds true when $v=0.2$, corresponding to half of the maximum velocity. We could, e.g., require that when the sensor measurements are not received one in three control cycles, the robot show reduce its translational speed to $v=0.2$. Such specification can be written as:
\begin{equation}
\label{LTLFault2}
\Box (fail_{3,1} \to \bigcirc\bigcirc\bigcirc\, slow) 
\end{equation}
where $fail_{3,1}$ captures one sensor failure in three control cycles:
$$(f \land \bigcirc f \land \bigcirc\bigcirc \lnot f) \lor (\lnot f \land \bigcirc f \land \bigcirc \bigcirc f) \lor (f \land \bigcirc \lnot f \land \bigcirc\bigcirc f).$$
By conjoining~(\ref{LTLFault}) with (\ref{LTLFault2}) we would obtain a more detailed requirement asking for the robot to slow down when one measurement fails in the three consecutive control cycles, and to stop when two measurements fail.

Table~\ref{results} show the time and space complexity of fine automata for formula
$\varphi = \Box (fail_{n,k} \to \bigcirc^n stop)$ where,
$k$ is the number of faults in $n$ consecutive readings.
The length column denotes the length of $nnf(\lnot \varphi)$.
$\bigcirc^n \phi$ is a shorthand of $n$-consecutive $\bigcirc$ applied to $\phi$.
\begin{figure}[htb]
\centering
\small{
\begin{tabular}{| l | r| r | r | r |}\hline
Parameters  &  $length$ &  Time(s) & |NFA| &  |DFA| \\
\hline
n =3, k=2 &  10 & 0.714 & 245 &  10\\
n=3, k=1   &  10 & 1.096   & 253 &  10 \\
n=4, k=1  &  13 & 12.690  & 1045 & 15 \\
n=5, k=1  &  16 & 110.026  & 2717 & 21 \\
n=6, k=1 &  19 & 1957.450  & 7933 & 28\\ 
\hline
\end{tabular}}
\caption{Fine Automata Size and Time to build}\label{results}
\end{figure}

\subsection{Mode-switching}
In this section we consider an instantiation of the mode-switching problem that frequently occurs in the autonomous vehicles. This  problem consists in defining different scenarios and specifying the desired behavior for each of those scenarios. In a cruise control system, for example, the nominal scenario would require maintaining a desired velocity. However, in the presence of rain or ice, the velocity may need to be reduced. Similarly, if the vehicle in front reduces its speed, an automatic cruise control system would immediately reduce the velocity to avoid a collision. Similar examples of scenarios and corresponding goals can be found in many different application domains.
To model the mode switching problem in LTL we consider first the template formula $\varphi_i$ defined as:
$$scen_i\implies (scen_i \land\neg goal_i) \mathsf{W} ((scen_i \land goal_i)\mathsf{W} \neg scen_i).$$
This formula is satisfied when if the scenario $i$ happens, then the system should stay in
scenario $i$ state until another scenario happens. Moreover,
when the syatem stays in the scenario $i$, it shoud try to reach $goal_i$ states.
%
%
If we have $n$ pairs of scenarios and goals, we can construct a formula $\varphi_i$ and the final requirement is captured by requiring the conjunction of these formulas to hold for all time:
$$\Box(\varphi_1\land \varphi_2\land\hdots\land \varphi_n).$$

\begin{figure}[htb]
\centering
\includegraphics[scale=0.42]{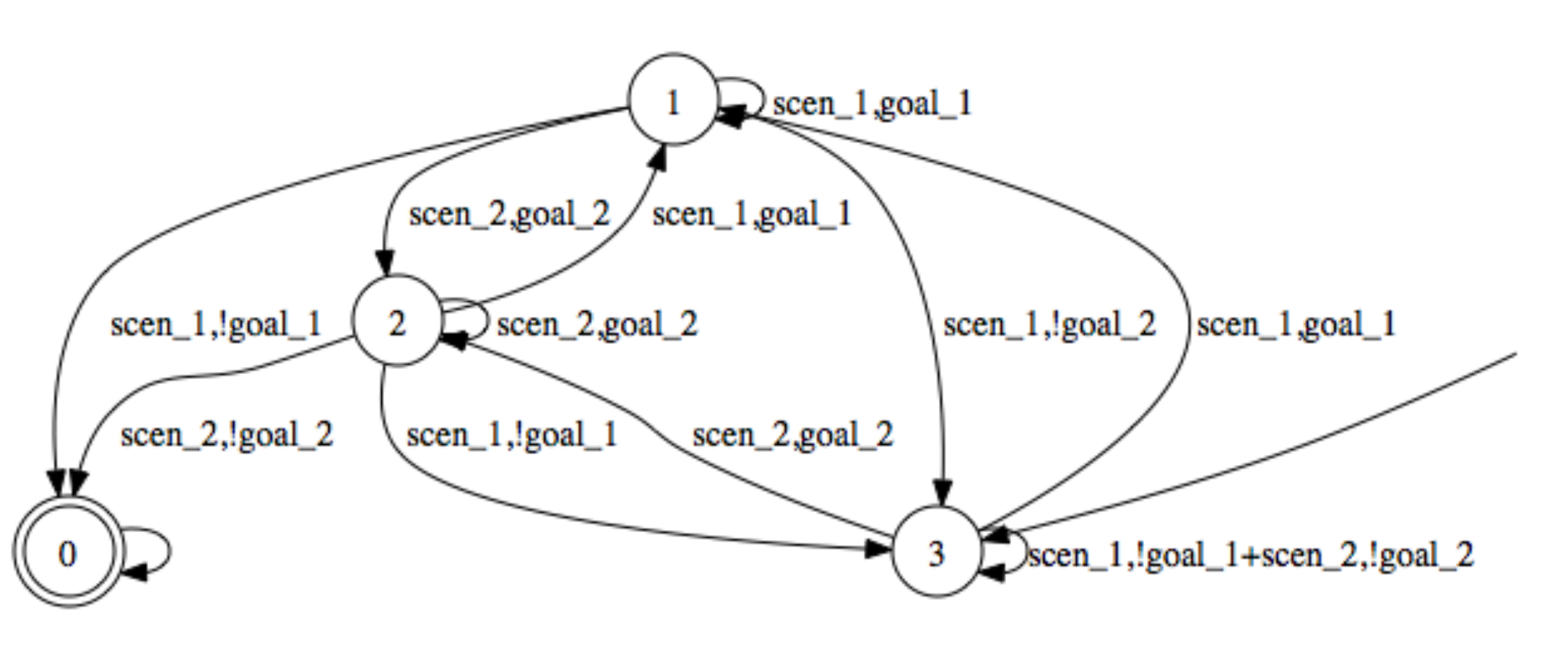}
\caption{Fine Automaton For Switching Property}\label{fig:switch}
\end{figure}

To illustrate the mode switching problem in the context of the mobile robot example, we consider the scenario to be specified by a remote operator that instructs the robot to move to one of two locations described by the predicates: 
$$goal_1=\{(x,y,\theta)\in \R^3\,\,\vert\,\, 4.4\le x\le 4.6\land 1\le y\le 1.6\}$$
$$goal_2=\{(x,y,\theta)\in \R^3\,\,\vert\,\, 4.6\le x\le 5.0\land 1\le y\le 1.6\}$$
The formulas defining the scenarios are the predicates $scen_1$ and $scen_2=\neg scen_1$ whose truth value can be dynamically changed by the robot operator according to the location where he wants the robot to go.
The fine automaton for the resulting specification (Figure~\ref{fig:switch}) was constructed in <1 seconds and has 4 dfa states. 
